\documentclass{article}
\usepackage{spconf,amsmath,graphicx}
\usepackage{booktabs}
\usepackage{amsmath}

\usepackage{amsmath,amssymb,amsfonts}
\usepackage{bm}
\usepackage{epsfig,graphicx,subfigure}
\usepackage{color}
\usepackage[normalem]{ulem}
\usepackage{multirow}
\usepackage{booktabs}

\usepackage[lowtilde]{url}
\usepackage[hidelinks]{hyperref}
\usepackage{cite}
\usepackage[ruled,vlined]{algorithm2e}
\usepackage{bm}
\usepackage{paralist}

\makeatletter
\def\blfootnote{\gdef\@thefnmark{}\@footnotetext}
\makeatother

\title{FeatherTTS: Robust and Efficient attention based Neural TTS}
\name{Qiao Tian$^1$, Zewang Zhang$^1$, Chao Liu$^{2*}$, Heng Lu$^1$, Linghui Chen$^1$, Bin Wei$^3$, Pujiang He$^3$, Shan Liu$^1$}

\address{
  $^1$Tencent,
  \\$^2$Harbin Institute of Technology(Shenzhen), 
  \\$^3$Intel Corporation\\
  {
  \small \tt \{briantian, zewangzhang, bearlu, nedchen\}@tencent.com 
  }
}

\begin{document}
\maketitle
\begin{abstract}
\label{sec:abs}
Attention based neural TTS is elegant speech synthesis pipeline and has shown a powerful ability to generate natural speech. However, it is still not robust enough to meet the stability requirements for industrial products.
Besides, it suffers from slow inference speed owning to the autoregressive generation process.
In this work, we propose FeatherTTS, a robust and efficient attention-based neural TTS system. Firstly, we propose a novel Gaussian attention which utilizes interpretability of Gaussian attention and the strict monotonic property in TTS. By this method, we replace the commonly used stop token prediction architecture with attentive stop prediction. Secondly, we apply block sparsity on the autoregressive decoder to speed up speech 
synthesis. The experimental results show that our proposed FeatherTTS not only nearly eliminates the problem of word skipping, repeating in particularly hard texts and keep the naturalness of generated speech, but also speeds up acoustic feature generation by 3.5 times over Tacotron.
Overall, the proposed FeatherTTS can be $35$x faster than real-time on a single CPU.
\end{abstract}

\blfootnote{*This work was done during internship in Tencent.}

%
%
\section{Introduction}
\label{sec:intro}


In recent years, with the rapid development of deep learning, neural text-to-speech (TTS) can synthesize speech which is more natural and expressive than traditional TTS pipeline.
Neural TTS is usually divided into two parts: an acoustic model and a neural vocoder. First, the input text (phoneme) sequence is converted into an intermediate acoustic feature sequence(linear spectrogram or mel-spectrogram) through an acoustic model such as Tacotron \cite{wang2017tacotron}, Tacotron2 \cite{shen2018natural}, Transformer TTS \cite{li2019neural}, FastSpeech \cite{ren2019fastspeech}, etc. Then, the Griffin-Lim algorithm \cite{griffin1984signal} or neural vocoder such as WaveNet \cite{van2016wavenet} and WaveRNN \cite{luong2015effective} is used to generate the final waveform according to the acoustic features.
Sequence-to-sequence models with an attention mechanism are currently the predominant paradigm in neural acoustic model and have shown a powerful ability to generate expressive and high-quality speech.
Those models learn the alignment between text sequence and frame-level acoustic features through the attention mechanism, and then predict spectral features that contain information such as pronunciation and prosody.
The speech quality synthesized by the neural TTS is limited by the alignment generated by the attention mechanism.
Although attention-based neural TTS has achieved great success, it is difficult to deploy in the industry due to its accidental alignment errors.

Tacotron \cite{wang2017tacotron} with content-based attention mechanism does not take into account the monotonicity and locality of TTS alignment, an improved hybrid location-sensitive mechanism proposed in Tacotron2 \cite{shen2018natural} combines content-based and location-based features to achieve the synthesis of longer utterances. However, such hybrid mechanism also causes alignment issues occasionally.
Recently, inspired by the purely location-based GMM attention mechanism\cite{graves2013generating}, an improved location-based GMM attention mechanism called GMMv2b is proposed in Google's work\cite{battenberg2020location}, which shows that the GMMv2b-based mechanism is able to generalize to long utterances, and can also improve speed and consistency of alignment during training.
However, the commonly used stop token architecture often causes early stop phenomenon for complex texts and long sentences.
In addition, such GMM attention is unnormalized and not strictly monotonic, which leads to unstable performance.

In this paper, we propose a novel attention-based neural TTS model named FeatherTTS, which can perform stable, fast and high-quality synthesis. Our major contributions are as follows:
(1) We introduce the Gaussian attention for acoustic modeling, a monotonic, normalized and stable attention mechanism, which is very interpretable for end to end speech synthesis.
(2) To solve the stop early issue, we remove the widely adopted stop token architecture in Tacotron2 and propose the attentive stop loss (ATL), which can determine whether to stop directly based on alignment and fast convergence for Gaussian attention.
(3) To improve the inference speed and reduce the number of parameters without sacrificing the speech quality, we propose to adopt block sparse strategy to prune the weights of decoder .

\section{Related work}
\label{sec:re}

\subsection{Hybrid attention based Tacotron2}
\label{sec:mwrnn}

Sequence-to-Sequence models with an attention mechanism  are currently the predominant paradigm in neural TTS. 
Attention-based neural TTS such as Tacotron2 \cite{shen2018natural} generally uses an encoder to encode input sequence $x_{1:J}$ into hidden representation $h_{1:J}$ as
\begin{equation}
    \{\bm{h}_{1:J}\} = Encoder(\{\bm{x}_{1:J}\}), \label{con:cbhgencoder}
\end{equation}
where $J$ is the length of input phoneme sequence. 
Then, the attention RNN generates a state vector $s_i$, which is used as the query vector of the attention mechanism to generate alignment $\alpha_i$ at decode time i.
According to the alignment $\alpha_i$, a weighted average of the encoder output is calculated, which is the context vector $c_i$.
\begin{equation}
    s_i = RNN_{Att}(s_{i-1}, c_{i-1}, y_{i-1})\label{con:attentionrnn}
\end{equation}
\begin{equation}
    \alpha_i = Attention(s_i, ...)\qquad c_i = \sum\limits_i\alpha_{i,j}h_j\label{con:attention}
\end{equation}

Finally, the context vector $c_i$ is fed into the decoder, and the final acoustic feature sequence ${y}_{1:T}$ is computed through post-net as
\begin{equation}
    d_i = {RNN}_{Dec}(d_{i-1}, c_i, s_i) \qquad y_i = f_o(d_i),\label{con:decoderrnn}
\end{equation}
where T is the length of output mel-spectrogram sequence.


Recently, many works have proposed various attention mechanism. Such as Tacotron~\cite{wang2017tacotron} uses the purely content-based attention mechanism introduced in~\cite{bahdanau2014neural}, Tacotron2\cite{shen2018natural} uses an improved hybrid location-sensitive mechanism introduced in \cite{chorowski2015attention}, 
some works \cite{raffel2017online,zhang2018forward, he2019robust} explore the use of monotonic attention mechanisms, and some authors \cite{kastner2019representation,battenberg2019effective} use the location-based GMM attention.

\subsection{Location based GMMv2b}
\label{sec:gmmv2b}

Recently, Google's work \cite{battenberg2020location} proposed a modified location-based attention mechanism which is called GMMv2b, has achieved great success. The GMMv2b mechanism is inspired by the location-based GMM attention mechanism introduced in \cite{graves2013generating}. The location-based GMM attention introduced in \cite{graves2013generating} uses K Gaussian components to compute the alignment $\alpha_i$ as (\ref{con:location-basedGMM}). 
The mean of each Gaussian component is computed following the recurrence relation in (\ref{con:mean}).
The monotonicity of GMM attention is guaranteed by making $\Delta_i$ non-negative. 
\begin{equation}
    \alpha_{i,j} = \sum^K_{k=1}{\frac{\omega_{i,k}}{Z_{i,k}}\exp{(-\frac{{(j-\mu_{i,k})}^2}{2{(\sigma_{i,k})}^2})}} \label{con:location-basedGMM}
\end{equation}
\begin{equation}
    \mu_i = \mu_{i-1} + \Delta_i \label{con:mean}.
\end{equation}

GMM attention usually calculates the intermediate variables ($\hat{\omega}_i$, $\hat{\Delta}_i$, $\hat{\sigma}_i$) first, and then uses the exponential function to obtain the final variables.
In order to stabilize GMM attention, GMMv2b-based attention uses the softmax and the softplus functions to compute the final mixture parameters as 
\begin{equation}
    \left\{\begin{array}{c}
        Z_i = \sqrt{2\pi\sigma^2_i} , \\
        \omega_i = S_{max}(\hat{\omega}_i) ,\\
        \Delta_i = S_+(\hat{\Delta}_i) ,\\
        \sigma_i = S_+(\hat{\sigma}_i) ,
    \end{array}
    \right.
\end{equation} 
where $S_{max}$ and $S_+$ are the softmax function and the softplus function respectively.
Besides, GMMv2b-based attention adds initial biases to the the intermediate parameters $\hat{\Delta}_i$ and $\hat{\sigma}_i$, which can encourage the final parameters to take on useful values at initialization.


As shown in \cite{battenberg2020location}, the GMMv2b-based mechanism is able to generalize to long utterances and maintains good naturalness, which makes the synthesis of the entire paragraph possible.

\section{The proposed method}
\label{sec:FeatherTTS}
\begin{figure}[ht]
\centering
	\begin{minipage}[b]{1.0\linewidth}
		\centerline
        \centerline{\includegraphics[width=8cm, height=6cm]{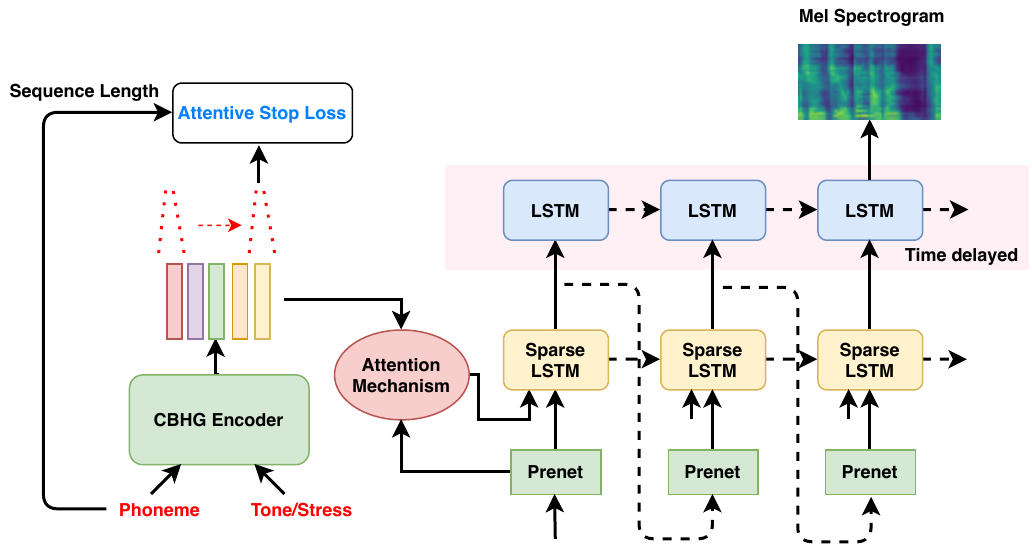}}
	\end{minipage}
	\caption{The architecture of FeatherTTS}
	\label{fig:fw_arc}
\end{figure}

Although the GMMv2b-based mechanism has good performance, it also has many problems.
First, this model still use stop token architecture which can lead to early stop.
Second, GMM attention isn't completely monotonic because it uses a mixture of distributions with infinite support.
Finally, GMMv2b attention is unnormalized because that attention weights are sampled from a continuous probability density function, this can lead to occasional spikes or dropouts in the alignment.
Especially, there are repetition problems for the synthesis of short sentences, such as monophone and vowel. Therefore, we propose FeatherTTS, a more robust attention-based acoustic model, as shown in Fig.~\ref{fig:fw_arc}. Our model is based on the Tacotron2~\cite{shen2018natural} architecture and consists of a CBHG encoder, Gaussian attention and a block sparse decoder.

\subsection{Gaussian attention}
\label{sec:at}
In order to solve the incomplete monotonic and unnormalized problem in GMM attention, we propose to use Gaussian attention mechanism to model alignment, as shown in (\ref{con:singleGaussian}). We also calculate the intermediate variables ($\hat{\sigma_i}$, $\hat{\Delta_i}$) first, and then get the final parameters($\sigma_i$, $\Delta_i$) through the softplus function.
\begin{equation}
   \alpha_{i,j} = \exp{(-\frac{{(j-\mu_{i})}^2}{2{(\sigma_{i})}^2})} \label{con:singleGaussian}
\end{equation}
\begin{equation}
    \mu_i = \mu_{i-1} + \Delta_i \label{con:singlemean}
\end{equation}

We use such simple and normalized Gaussian attention function to calculate the alignment $\alpha_{i,j}$. The mean $\mu_i$ and the variance $(\sigma_i)^2$ of the Gaussian attention mechanism control the position and width of the attention window, respectively. 
$\Delta_i$ is non-negative, so the mean $\mu_i$ is monotonically increasing, which guarantees the alignment process of the Gaussian attention mechanism is completely monotonic.


\subsection{Attentive stop loss}
\label{sec:atl}
The stop token architecture used in Tacotron2 \cite{shen2018natural} will cause stop early problems.
In addition, the alignment information learned in the Gaussian attention is too weak, which makes the alignment difficult to converge.
In order to solve the above problems, we remove the stop token architecture, and propose the attentive stop loss, which directly judges the stop based on alignment. It is calculated as
\begin{equation}
    L_{stop} = \left| \mu_{T} - (J+1)\right|, \label{con:stoploss}
\end{equation}
where $\mu_T$ is the mean value of Gaussian attention function at last step, and $J$ is the length of input phoneme sequence.

During training, the attentive stop loss forces the mean $\mu_i$ of Gaussian attention to go forward to the end of the phoneme sequence to ensure accurate alignment.
In the inference stage, FeatherTTS will stop to predict when $\mu_i \geq (J+1)$.




\subsection{Sparse autoregressive decoder}
\label{sec:sd}

It has been demonstrated that, with the same computational complexity, a larger sparse network behaves better than a smaller dense network~\cite{luong2015effective,valin2019lpcnet}. In this work, to reduce the amount of computation of LSTM layers in decoder without a significant loss in quality, we reduce the number of non-zero values in each LSTM kernel weight. Inspired by~\cite{narang2017block,narang2017exploring}, we adopt the weight pruning scheme based on the weight magnitude. 

We start to perform weight pruning after 20K steps and every 500 steps, we sort the weights of sparsified LSTM layers and zero out certain number of weights with the smallest magnitudes until the target sparsity $90\%$ is reached at 200K step. After block sparsity, the number of main operations in every sparsified LSTM layer is 
\begin{equation}
    C = 4\textsl{(1 - S)} (I*H + H^2) \label{con:sparseLSTM},
\end{equation}
where $I$ and $H$ are the dimensions of input and hidden state of the LSTM cell, respectively, and $S$ is the target sparsity.  

In FeatherTTS, we used the time-delayed post-net as in \cite{zhang2020adadurian}, which is a vanilla LSTM layer with 256 units.
Overall, FeatherTTS is trained to minimize the total loss as 
\begin{equation}
    Loss= \frac{1}{T}\sum^T_{i=1}\left| y'_i - {y}_i\right| + \frac{1}{T-d}\sum^{T-d}_{i=1}\left| y''_{i+d} - {y}_i\right| + \lambda L_{stop}, \label{con:totalloss}
\end{equation}
where d is the number of frames of time delay and $\lambda$ is a scaling factor. On the right hand side of Eq.~\ref{con:totalloss}, the first two items of the loss function are L1 loss between reference mel-spectrogram ${y}_i$ and the predicted both before and after mel-spectrogram $y'_i$, $y''_i$. The last item is the attentive stop loss.

\section{Experiments}
\label{sec:exps}

\subsection{Data Set}
\label{ssec:data}
We used a corpus containing 20 hours of Mandarin corpus recordings by a professional broadcaster for all experiments. The corpus was split into a training set of approximately 18 hours and a test set of 2 hours. All the recordings were down-sampled to 24KHz sampling rate with 16-bit format. We used 80-band mel-scale spectrogram as training target, and then the mel-scale spectrogram was converted into waveforms by FeatherWave neural vocoder\cite{tian2020featherwave}.
 
\subsection{Experimental Setup}
\label{ssec:vocoders}
For comparisons, we implemented two models including GMMv2b-based Tacotron2 and FeatherTTS. As the baseline model, the GMMv2b-based model is composed of five mixture components.
In order to reduce the model size, training and inference time, two consecutive frames were predicted at each decoding time step. 
For FeatherTTS, we delayed $5$ frames and the rate of attentive stop loss $\lambda$ was set to $0.001$.
All models were trained 300k steps with batch size 32 on a single GPU.
Other experimental setups are the same as AdaDurIAN \cite{zhang2020adadurian} if not specified.

\subsection{Evaluations}
\label{ssec:mos}
In this section, we evaluated the proposed FeatherTTS and Tacotron2 (GMMv2b) in term of naturalness and robustness, and compared the synthesis speed of the above two models and FastSpeech.
\subsubsection{Mean Opinion Score}
We used the Mean Opinion Score (MOS) to measure the naturalness of the synthesized speech\footnote{Part of synthesized samples could be found at this URL:\\\url{https://wavecoder.github.io/FeatherTTS/}}.
We used 20 unseen sentences for evaluating the models. MOS of the naturalness of generated utterances rated by human subjects participated in the learning tests through crowdsourcing. The results of subjective MOS evaluation are presented in Table \ref{table:mos}. 
The results show that, under the same vocoder configuration, both FeatherTTS and Tacotron2(GMMv2b) have similar MOS values. 
In addition, we compared the effect of block sparsity on the sound quality. It can be seen from the experimental results that FeatherTTS with block sparsity outperforms FeatherTTS without block sparsity with a gap of 0.01 in MOS, which is basically in line with our expectations.

\begin{table}
\centering
\caption{Mean Opinion Score (MOS) with $95\%$ confidence intervals for different models.}
\begin{tabular}{cc}
\toprule
{\textbf{Model}} & {\textbf{MOS on speech quality}} \\
\midrule
Tacotron2(GMMv2b) & 4.31 $\pm$ 0.03 \\
FeatherTTS w/o Block sparsity & 4.32 $\pm$ 0.04 \\
\textbf{FeatherTTS} & \textbf{4.33 $\pm$ 0.04} \\
\bottomrule
\end{tabular}
\label{table:mos}
\end{table}

\subsubsection{Word Error Rate}
The design goal of FeatherTTS is to keep the naturalness as Tacotron2(GMMv2b) while avoiding the mispronunciations observed in the Tacotron2(GMMv2b). Therefore, we compared the robustness of two systems in terms of generated speech. To evaluate the robustness of FeatherTTS, we prepared 20 hard sentences for two systems and focused on the word skipping, word repetition and inaccurate intonation. The results are shown in Table \ref{table:wer}. 
We can see that Tacotron2(GMMv2b) has an error rate of $4.1\%$, while FeatherTTS is more robust, with an error rate of only $0.9\%$. 
This strongly proves the role of Gaussian attention and attentive stop loss in improving model stability.

\begin{table}
\centering
\caption{The Word Error Rate (WER) for different models.}
\begin{tabular}{cc}
\toprule
{\textbf{Model}} & {\textbf{Word error rate}} \\
\midrule
Tacotron2(GMMv2b) & 4.1\% \\
\textbf{FeatherTTS} & \textbf{0.9\%} \\
\bottomrule
\end{tabular}
\label{table:wer}
\end{table}


\subsubsection{Synthesis Speed}
\label{ssec:speed}
In this experiment, we proved the effectiveness of the proposed block sparse decoder for accelerating training and inference. We compared the real-time rate of FastSpeech, Tacotron2(GMMv2b) and FeatherTTS to generate mel-spectrograms on a single core CPU(Intel Xeon Platinum 8255C). The results of synthesis speed are presented in Table \ref{table:com}. 
Tacotron2(GMMv2b) can achieve an inference speed of 3.5 times faster than real time on CPU with single core, while FeatherTTS can further be accelerated by 3.5 times over Tacotron2(GMMv2b). In addition, compared with non-autoregressive FastSpeech, FeatherTTS is also about 2.6 times faster .
Furthermore, we truncated the parameters and ran them on the BF16 format to reduce the memory consumption, and finally achieve 60 times faster than real-time on a single CPU core (Cooper Lake, 3rd Gen Intel Xeon Scalable processors).
The above experiments prove the accelerating performance of the block sparse decoder for inference, and makes it possible to deploy TTS on edge devices. 


\begin{table}
\centering
\caption{The inference speed of different models.}
\begin{tabular}{ccc}
\toprule
{\textbf{Model}} & {\textbf{Speed}} \\
\midrule
FastSpeech  & 13.3x \\
Tacotron2(GMMv2b) & 10.4x \\
\textbf{FeatherTTS} & \textbf{35.0x} \\
\textbf{FeatherTTS {\textbf{BF16}}} & \textbf{60.0x} \\
\bottomrule
\end{tabular}
\label{table:com}
\end{table}

\section{Conclusions}
\label{sec:cons}
In this work, we proposed FeatherTTS, an improved neural TTS system with Gaussian attention, attentive stop loss and block sparse decoder. 
Experiments demonstrate that such attention mechanism is very efficient and would greatly improve robustness of attention-based neural TTS system. 
With block sparse decoder, our proposed FeatherTTS can speed up the mel-spectrogram generation by 3.5 times faster than Tacotron2 nearly without any performance degradation.
The ideas introduced in FeatherTTS pave a new way for both efficient and robust speech synthesis, and could be also applied to other sequence-to-sequence task including automatic speech recognition.

For future work, we will continue to investigate the performance of FeatherTTS on edge-devices.
\section{Acknowledgments}
\label{sec:ack}
The authors would like to thank Yi Xie in IAGS, Intel Asia-Pacific Research $\&$ Development Co Ltd..  This member in Intel helped to optimized our algorithm with AVX512 and BF16 intrinsics to get good performance on the 3rd Gen Intel Xeon Scalable processors.

\vfill\pagebreak

\bibliographystyle{IEEEbib}
\bibliography{strings,refs}
\end{document}